\documentclass[iop,revtex4]{emulateapj}    

\usepackage{natbib}
\usepackage{amsmath}
\usepackage{color}
\usepackage{longtable}
\usepackage{graphicx}
\usepackage{lscape}
\usepackage{textcomp}

\usepackage{hyperref}

\newcommand{\kms}{\,km\,s$^{-1}$}

\newcommand{\mum}{\,$\mu$m}

\newcommand{\Msun}{$M_{\odot}$}

\newcommand{\Msunyr}{$M_{\odot}$yr$^{-1}$}

\newcommand{\rev}{ }

\begin{document}

\title{A  candidate planetary-mass object with a photoevaporating disk in Orion}
\author{Min~Fang\altaffilmark{1}, Jinyoung~Serena~Kim\altaffilmark{1}, Ilaria~Pascucci\altaffilmark{2}, D\'aniel~Apai\altaffilmark{1, 2}, Carlo~Felice~Manara\altaffilmark{3}}
\altaffiltext{1}{Department of Astronomy, University of Arizona, 933 North Cherry Avenue, Tucson, AZ 85721, USA}
\altaffiltext{2}{Department of Planetary Sciences, University of Arizona, 1629 East University Boulevard, Tucson, AZ 85721, USA}
\altaffiltext{3}{Scientific Support Office, Directorate of Science, European Space Research and Technology Centre (ESA/ESTEC), Keplerlaan 1,
2201 AZ Noordwijk, The Netherlands}

\begin{abstract} 
 In this work, we report the discovery of a candidate planetary-mass object with a photoevaporating protoplanetary disk, Proplyd~133-353, which is near the massive star $\theta^{1}$\,Ori\,C at the center of the Orion Nebula Cluster (ONC). The object was known to have extended emission pointing away from  $\theta^{1}$\,Ori~C, indicating ongoing external photoevaporation. Our near-infrared spectroscopic data {\rev and the location on the H-R diagram} suggests that the central source of Proplyd~133-353 is substellar ($\sim$M9.5), and have a mass probably less than 13 Jupiter mass and an age younger than 0.5\,Myr. Proplyd~133-353 shows a similar ratio of X-ray luminosity to stellar luminosity to other young stars in the ONC  with a similar stellar luminosity, and has a similar proper motion to the mean one of confirmed ONC members. We propose that Proplyd~133-353 formed in a very low-mass dusty cloud {\rev or an evaporating gas globule} near $\theta^{1}$\,Ori\,C as a second-generation of star formation, which can explain both its young age and the presence of its disk.

\end{abstract}

\maketitle

\section{Introduction}
Protoplanetary disks in massive clusters can be rapidly dissipated \citep{2007ApJ...660.1532B,2012A&A...539A.119F,2016arXiv160501773G}. The main physical mechanism driving disk dispersal in these harsh environments is thought to be photoevaporation driven by UV photons from the massive stars in the clusters \citep{1998ApJ...499..758J,1999ApJ...515..669S,2000ApJ...539..258R,2007MNRAS.376.1350C,2013ApJ...774....9A,2016MNRAS.457.3593F}. UV photons from massive stars ionize and heat the gas in the disk surface and induce a gas flow away from the disk when the sound speed of the gas exceeds the escape velocity \citep{1998ApJ...499..758J,2000prpl.conf..401H}. Direct evidence for this mechanism is provided by the large numbers of proplyds found near massive stars \citep{1993ApJ...410..696O,2012ApJ...746L..21W,2012A&A...539A.119F,2016ApJ...826L..15K}. The proplyds have cometary structures with tails pointing away from the nearby massive stars and are interpreted as the outer regions of disks of young stars that are being photoevaporated by extreme ultraviolet (EUV) and far ultraviolet (FUV) radiation from the massive stars \citep{1993ApJ...410..696O}. 

Until now, the largest sample of proplyds has been found to be around $\theta^{1}$\,Ori\,C, which is an O6-type massive star in the Orion Nebula cluster (ONC) at a distance of 414\,pc \citep{2007A&A...474..515M}. A complete sample of such proplyds has been cataloged in \citet{2008AJ....136.2136R}. For our investigation, with the MMT and Magellan infrared spectrograph \citep[MMIRS][]{2012PASP..124.1318M} mounted on the MMT telescope, we performed near-infrared spectroscopic observations of the central stars in a subsample of the proplyds (Fang~et~al.~in~preparation). 

In this work, we present a study of an extremely interesting case, Proplyd~133-353 in \citet{2008AJ....136.2136R}.  Based on our spectroscopy, the central object of the proplyd has a very late spectral type ($\sim$M9.5), which suggests that it has a mass in the planetary range at the age of the ONC \citep[1\,Myr][]{1997AJ....113.1733H}. We have organized paper as follows: in $\S$2 we briefly introduce Proplyd 133-353, in $\S$3 we describe our observations,  data reduction, and delineate our data analysis, in  $\S$4, we  present our results and  discussion, followed by a summary in $\S$5.

\begin{figure}
\begin{center}
\includegraphics[width=1\columnwidth]{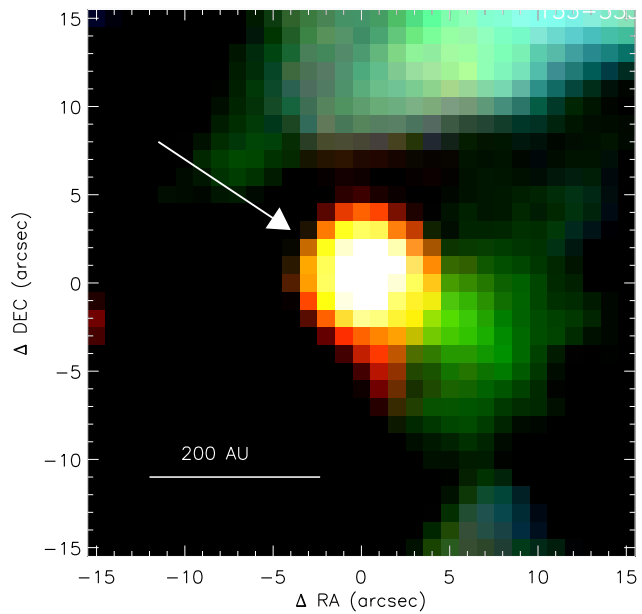}
\caption{HST ACS color-composite image of Proplyd 133-353 (blue: F555W, green: F658N, red: F850LP). The tail of Proplyd 133-353 is obvious in the HST image in the F658N band. The arrow  shows the direction from $\theta^{1}$\,Ori\,C to Proplyd 133-353.}\label{Fig:proplyd}
\end{center}
\end{figure}

\section{Proplyd 133-353}\label{Sect:Proplyd}
Proplyd~133-353 at RA(J2000.0)=$05^{\rm h}35^{\rm m}13.306^{\rm s}$, DEC(J2000.0)=$-$05$^{\circ}$23$^{'}$52.99$^{''}$  is located southwest of $\theta^{1}$\,Ori\,C. The projected distance between them is about 0.11\,pc. In Fig.~\ref{Fig:proplyd}, we show the HST ACS color-composite image of Proplyd 133-353 using the HST images from \citet{2013ApJS..207...10R}. The tail can be clearly seen in the HST image in the F658N band (green emission) and extends out to 190 AU. We visually inspected the HST images of the object in different bands from \citet{2013ApJS..207...10R}, and found that the tail can also be clearly detected in the F435W band for ACS/WFC, and the F336W and F656N bands for WFPC2, which could be due to the tail showing strong Balmer emission lines, e.g., H$\alpha$ (F658N), H$\gamma$ (F435W), and Balmer continuum emission (F336W). 

\begin{figure*}
\begin{center}
\includegraphics[width=2\columnwidth]{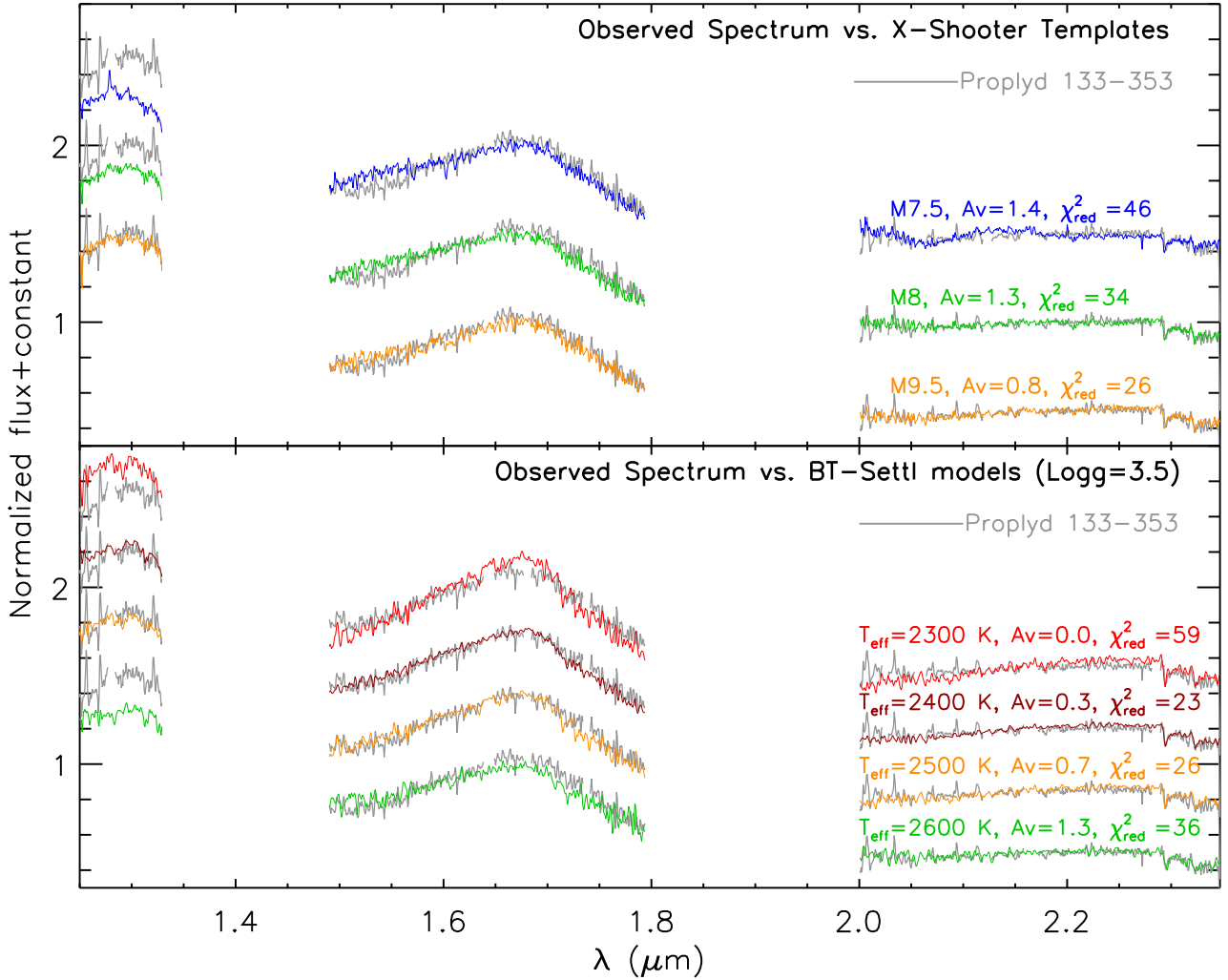}
\caption{MMIRS spectrum of Proplyd~133-353 (gray lines). The reddened X-Shooter templates (top) and the BT-Settl atmospheric models are shown for comparison. For  individual templates or models, their spectral types or effective temperature, as well as the best-fit visual extinctions and the $\chi_{\rm red}^{2}$ are marked.}\label{Fig:Spec}
\end{center}
\end{figure*}

\section{Data and analysis}
\subsection{Near-infrared spectroscopic data}
We observed the central object of Proplyd 133-353 during the night of 2015~Dec~31 in the MOS mode with MMIRS, as part of our spectroscopic survey of young stars in Orion. We used the HK grism and the HK3 filter, and created a mask with 0.5$''\times7''$ slits, following the MMT mask preparation procedure. This setup yields spectra from 1.25 to 2.34\,\mum, with an average spectral resolution of $\lambda/\Delta\lambda\approx$1100. We took 8 exposures (8$\times$300~second) for 4 spatially dithering pairs. After the scientific exposures, we observed one telluric standard HD~34481 at a similar airmass (airmass differences$<$0.02) to our scientific targets.

We reduced the data using the CFA MMIRS pipeline from \cite{2015PASP..127..406C}. The pipeline can reduce the MMIRS spectroscopic data in a fully automatic way. It subtracts pairs of dithered spectral exposures, performs flat-field and residual sky corrections, and rectifies the 2D spectra. The spectra are wavelength calibrated using airglow OH lines, and corrected for the telluric absorption using telluric standard stars we observed. A detailed description of the pipeline can be found in \citet{2015PASP..127..406C}.

\begin{figure*}
\begin{center}
\includegraphics[width=1\columnwidth]{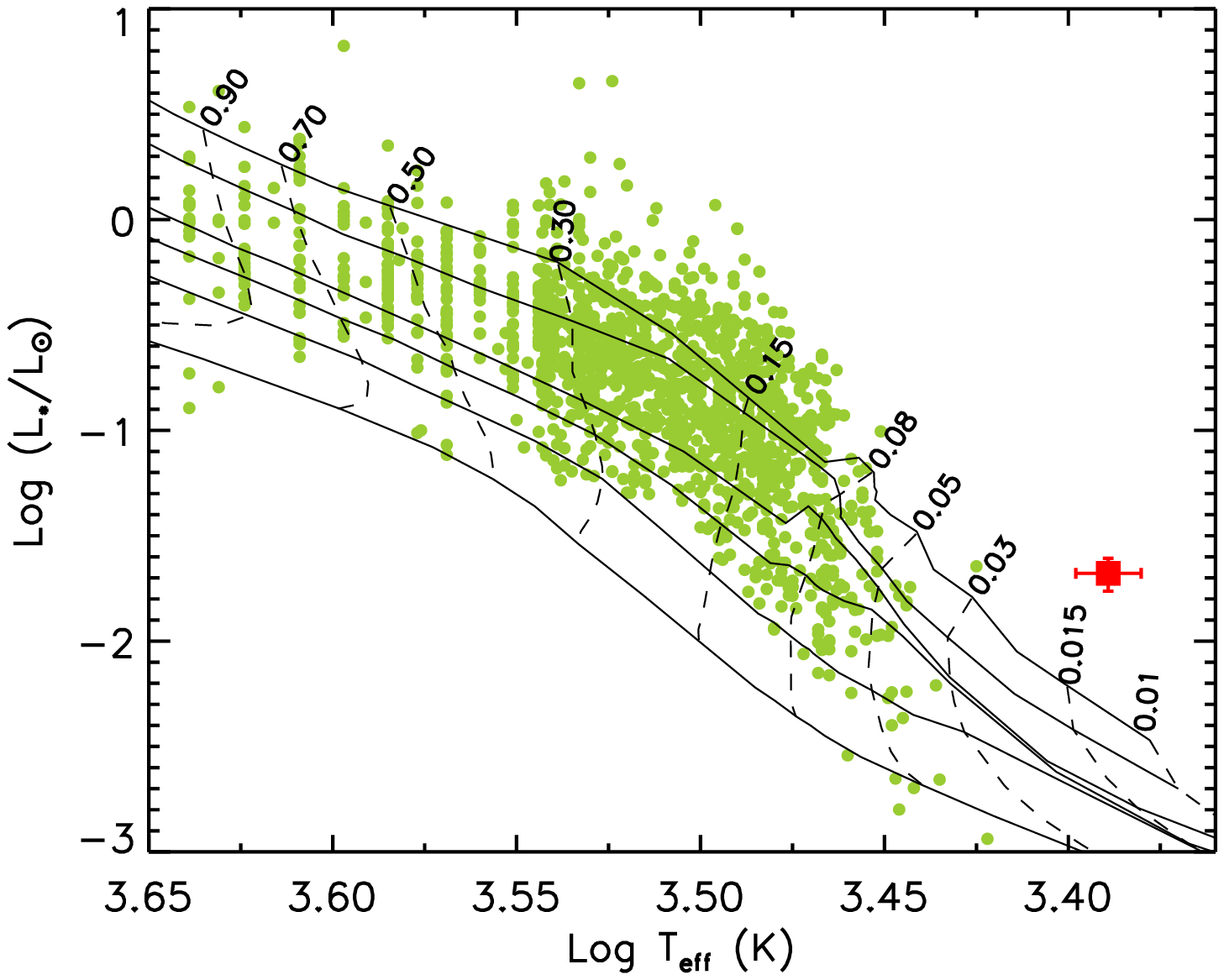}
\includegraphics[width=1\columnwidth]{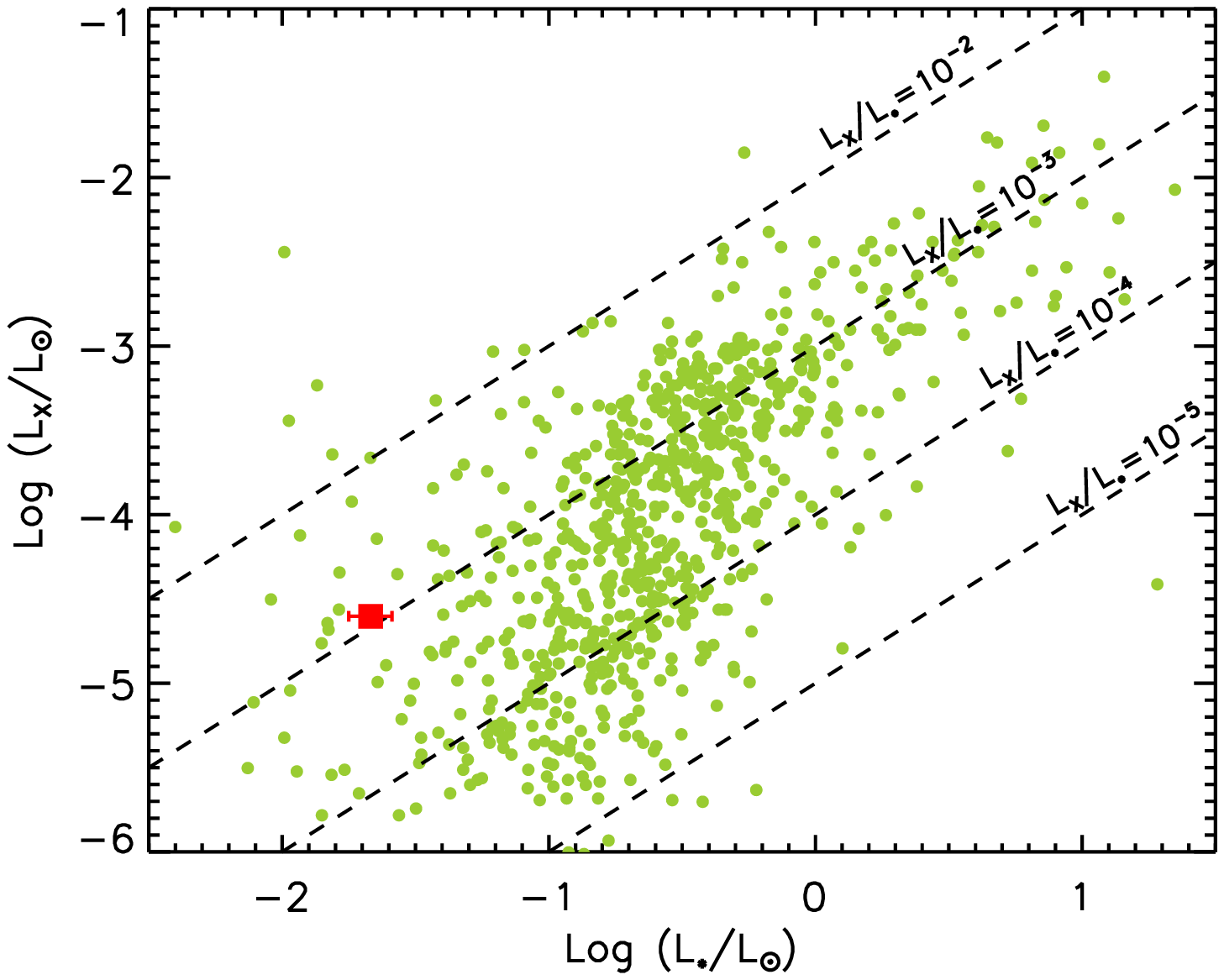}
\caption{Left: H-R diagram for young stars in the ONC. The filled box is for Proplyd~133-353, and the filled circles show other young stars in the ONC. The solid lines show the isochrones at ages of 0.5, 1, 3, 5, 10, and 30\,Myr, respectively, from \citet{2015A&A...577A..42B}. The dashed lines present the evolutionary tracks of young stellar/substellar objects with masses marked in the panel. Right: X-ray luminosity vs. stellar luminosity for young stars in the ONC. The filled box is for Proplyd~133-353, and the filled circles show other young stars in the ONC. The dashed lines mark $L_{\rm X}/L_{*}$ ratios of 10$^{-2}$, 10$^{-3}$, 10$^{-4}$, and 10$^{-5}$.}\label{Fig:HRD}
\end{center}
\end{figure*}

\subsection{Spectral classification}
We constructed a set of spectral templates with spectral types ranging from G5 to M9.5, using the spectra of diskless young stars collected with the instrument X-Shooter mounted on the Very Large Telescope \citep[][Manara et al. in preparation]{2013A&A...551A.107M}. Taking the visual extinction as a free parameter, we fit the observed spectra using the spectral templates which are reddened with the extinction law from \citet{1989ApJ...345..245C}, adopting a total to selective extinction value $R_{\rm V}=3.1$.  The {\rev best fit} is determined by minimizing $\chi_{\rm red}^{2}=\frac{1}{N-1}\sum_{i=1}^{N}\frac{(F_{\rm \lambda, obs}-F_{\rm \lambda, template})^{2}}{\sigma_{\rm \lambda, obs}^{2}}$, where $N$ are the numbers of the wavelength bins, $F_{\rm \lambda, obs}$ and $F_{\rm \lambda, template}$ are the observed fluxes and the ones from the spectral templates, respectively, within the wavelength $\lambda$ bin, and  $\sigma_{\rm \lambda, obs}$ are the errors in the observed fluxes. In Fig.~\ref{Fig:Spec}, we show a comparison of the observed spectrum of Proplyd~133-353 and the X-Shooter templates reddened with the best-fit visual extinction corresponding to the individual templates. We note that the template with spectral type M9.5 gives the best fit to the observation. However, a conversion of the spectral type M9.5 to the effective temperature ($T_{\rm eff}$) is still uncertain, and could range from  less than 2400\,K to $\sim$2500\,K using the different temperature scales in the literature \citep{2003ApJ...593.1093L,2013A&A...556A..15R,2014ApJ...786...97H}. To avoid this uncertainty, we fit the spectrum of Proplyd~133-353 directly with the BT-Settl atmospheric models with solar abundances from \cite{2009ARA&A..47..481A}, and a surface gravity Log~{\it g}=3.5, which is suitable for the very low-mass young substellar objects \citep{2015A&A...577A..42B}. As with our procedure for the X-Shooter templates, we took the visual extinction as a free parameter, and searched for the best-fit atmospheric models. Figure~\ref{Fig:Spec} shows a comparison of the observed spectrum and the reddened BT-Settl models. We found the best-fit BT-Settl models are the ones with the effective temperatures 2400 and 2500\,K with the best-fit visual extinction of 0.3 and 0.7\,mag, respectively. Thus, in this work, we adopt $T_{\rm eff}$=2450$\pm$50\,K and $A_{\rm V}=0.5\pm0.2$ for the central object of Proplyd~133-353.

\section{Results and Discussion}
\subsection{Masses and ages}
We obtained the near-infrared photometry of Proplyd~133-353 from \citet{2010AJ....139..950R}. The stellar luminosity of Proplyd~133-353 was derived using dereddened $J$-band photometry, and the $J$-band  bolometric correction ($BC_{J}$=2.12) for $T_{\rm eff}$=2450$\pm$50\,K calculated with the BT-Settl models, assuming a distance of 414\,pc \citep{2007A&A...474..515M}. The stellar luminosity ($L_{\star}$) of Proplyd~133-353 is estimated to be 0.021$\pm$0.004\,$L_{\odot}$. The uncertainty of $L_{\star}$ is calculated considering the uncertainty in the J-band magnitude, the $T_{\rm eff}$ value, and the extinction. In Fig.~\ref{Fig:HRD} (left), we placed Proplyd~133-353 in the H-R diagram. As a comparison, we also show other young stars in the ONC with data collected from \citet{2012ApJ...748...14D}, as well as the pre-main sequence (PMS) evolutionary tracks from \citet{2015A&A...577A..42B}. In the H-R diagram, Proplyd~133-353 is above the youngest PMS isochrone ($\sim$0.5\,Myr) from \citet{2015A&A...577A..42B}, and has a much cooler  $T_{\rm eff}$ compared with other young stars, but shows a  $L_{\star}$ similar to a 0.05\,\Msun brown dwarf at 1\,Myr.  {\rev Recently, \citet{2016arXiv160904041K} found the mean distance of the ONC to be 388$\pm$5\,pc. Using this distance estimate, the  $L_{\star}$ of Proplyd~133-353  will be reduced by a factor of 1.14, which moves the object down on the H-R diagram very slightly. Therefore, the object is still very young (age$<$0.5\,Myr) even with this new distance estimate.}

Due to the lack of PMS evolutionary tracks at ages younger than 0.5\,Myr in \citet{2015A&A...577A..42B},  it is not possible to derive its mass and age directly through comparison to the PMS evolutionary tracks. Here, we tentatively constrain its mass using the 0.5\,Myr isochrone from \citet{2015A&A...577A..42B}. On this isochrone, the mass of an object with  $T_{\rm eff}$=2450 is $\sim$13~Jupiter~mass ($M_{\rm J}$), {\rev which is much lower than the known lowest mass (42\,$M_{\rm J}$) for the objects at the centers of proplyds in ONC in the literature \citep{2008ApJ...687L..83R}}. However, this mass might be an upper limit for Proplyd~133-353, since the $T_{\rm eff}$ values of young low-mass substellar objects seem to decrease during their evolution.  The dividing line between a brown dwarf and a planet is around 13~$M_{\rm J}$, as defined by the deuterium fusion mass limit \citep{1997ApJ...491..856B,2011ApJ...727...57S}. Therefore, Proplyd~133-353 could be a planetary-mass object with an age younger than 0.5\,Myr. We also use the PMS evolutionary tracks from \citet{1997MmSAI..68..807D}, which gives a mass $\sim$25~$M_{\rm J}$ and an age $\sim$4000~yr. We note that evolutionary models at these young ages and very low masses have significant uncertainties \citep{2005Natur.433..286C,2010ApJ...711.1087K}, therefore in this work we do not attempt to conclusively determine the nature of the object. Whether in the brown dwarf or planetary mass category, Proplyd~133-353 is an interesting object for a {\rev detailed} study.

\subsection{Is Proplyd 133-353 a member of the ONC?}
As described in Section.~\ref{Sect:Proplyd}, Proplyd~133-353 has a tail pointing away from $\theta^{1}$\,Ori\,C, which indicates that its disk is being dissipated by the massive star. Thus, Proplyd 133-353 should be a member of the ONC. Here, we will discuss further evidence for the ONC membership of Proplyd~133-353  based on its X-ray properties and proper motion.

Proplyd~133-353 has been detected by the Chandra space telescope in Chandra Orion Ultradeep Project \citep{2005ApJS..160..319G}. In Fig.~\ref{Fig:HRD} (right), we show its X-ray luminosity ($L_{\rm X}$) over 0.5--8.0 keV band vs. its stellar luminosity, and compare it with the values from other young stars in the ONC with the $L_{\rm X}$  from \citet{2005ApJS..160..319G}, and $L_{stars}$  from \citet{2012ApJ...748...14D}. The $L_{\rm X}$/$L_{star}$ ratio of Proplyd~133-353 is similar to those of other young stars with similar stellar luminosity. 

\begin{figure}
\begin{center}
\includegraphics[width=1\columnwidth]{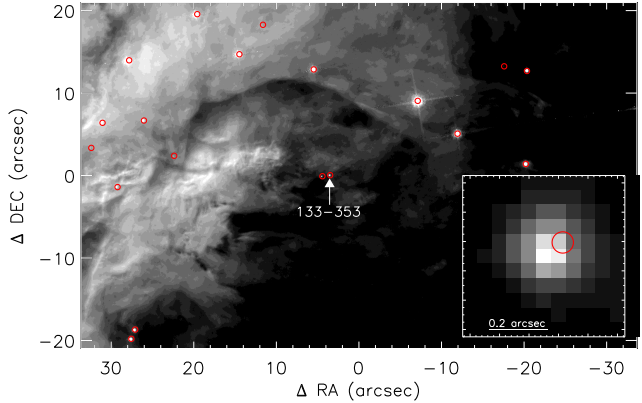}
\caption{HST F775W-band image of the field around Proplyd 133-353 observed in Feb, 2015. The arrow marks the position of Proplyd 133-353. The open circles show the positions of the stars on the HST F775W-band image observed in Apr, 2005. The inset zooms in on Proplyd 133-353.}\label{Fig:proper}
\end{center}
\end{figure}

According to the PMS evolutionary models {\rev \citep{2015A&A...577A..42B}}, the expected $L_{\star}$ for Proplyd~133-353 at an age of 1\,Myr is 0.00307\,$L_{\odot}$, corresponding to its $T_{\rm eff}$. If Proplyd~133-353 is a foreground 1\,Myr old planet, the expected distance could be $\sim$160\,pc, corresponding to the intrinsic luminosity and the observed brightness. If that is the case, we would expect that Proplyd~133-353 shows a different proper motion from the ones of other young stars in the ONC. We derive the proper motion from the HST archive images in F775W band observed at two epochs (2005~Apr.  vs. 2015~Feb.)\footnote{The first-epoch HST image is obtained from \citet{2013ApJS..207...10R}, and the second-epoch HST image (PropID~13826) is download from \href{url}{http://hla.stsci.edu/}}. In Fig.~\ref{Fig:proper}, we show the second-epoch HST image. The image has been calibrated in astrometry to the first-epoch image using the common ONC members. We excluded Proplyd~133-353 in the astrometric calibration; therefore we can measure the proper motion of the proplyd relative to the ONC. As a comparison, in Fig.~\ref{Fig:proper} we also show the positions of the stars in the first-epoch HST image. From the image pair, we obtain a relative proper motion for Proplyd~133-353, of $\mu_{\alpha}$=4$\pm$1~mas/yr, and $\mu_{\delta}$=$-$3$\pm$1~mas/yr. Therefore, there is no significant proper motion of Proplyd~133-353 relative to the ONC.  However, we cannot exclude the possibility that Proplyd~133-353 and the ONC have different motions in the line of sight, given the spectral resolution ($\lambda/\Delta\lambda\approx$1100) of our MMIRS data.

In conclusion, the arguments presented above, especially the cometary structure with a tail pointing away from $\theta^{1}$\,Ori\,C and the X-ray luminosity, strongly support that Proplyd~133-353 is a member of the ONC.

\subsection{The Possible Nature of  Proplyd~133-353}
 In the following we will compare the different scenarios for the origin and nature of Proplyd~133-353.

Theoretical calculations predict that mass loss rates from disks due to photoevaporation can be on the order of 10$^{-7}$\,\Msunyr \ within a distance of 0.2\,pc from the ionizing massive star $\theta^{1}$\,Ori\,C \citep{1999ApJ...515..669S}, which is confirmed by spectroscopic observations \citep{1999AJ....118.2350H}. The photoevaporation process can effectively dissipate protoplanetary disks outside the gravitation radii ($r_{\rm g}$), where the escape velocities equals the speed of sound, which is in turn determined by the UV heating \citep{2000prpl.conf..401H}. With such high  mass loss rates, we would expect a very short disk lifetime for Proplyd~133-353, which is $\sim10^{3}$\,yr with an assumption of a typical ratio ($\sim$100) between stellar mass and disk mass.

 It is well known that there are many tiny dusty clouds, called ``globulettes,'' around H\,{\scriptsize II} regions \citep{2006AJ....131.2580D,2007AJ....133.1795G,2014A&A...565A.107G}. The mean mass of the globulettes is around several$\times$10\,$M_{\rm J}$, and their mean radius is $\sim$4000\,AU \citep{2007AJ....133.1795G}. These globulettes are undergoing photoevaporation by UV photons from the massive stars in  H\,{\scriptsize II} regions. However, the shocks generated by photoionization, as well as the surrounding warm gas, can also exert external pressure on the globulettes, and drive their collapse to form brown dwarfs and free-floating planets before they are eroded by photoevaporation \citep{2007AJ....133.1795G}. Such a second-generation formation of  brown dwarfs and free-floating planets has been confirmed by the discovery of dense cores in some of the globulettes \citep{2013A&A...555A..57G}. \citet{2015MNRAS.446.1098H} doubt the possibility that the external pressure in the form of radiation or ram pressure can drive globulettes to form brown dwarfs or free-floating planets, and instead they  propose the collapse of the globulettes might be triggered by their collisions with {\rev the shells that borders the expanding H\,{\scriptsize II} regions}. If Proplyd~133-353 was formed in such a  globulette near $\theta^{1}$\,Ori\,C due to either  mechanism, we might explain both its younger age (with respect to the one of the ONC), and the lifetime of its disk, since its parental cloud can protect it from direct photoevaporation for a timescale of 10$^{5}$--$10^{6}$\,yr \citep{2007AJ....133.1795G,2015MNRAS.446.1098H}.


{\rev Finally, we explore an alternative scenario for the formation of Proplyd~133-353: Instead of a photo-evaporating disk it may be an evaporating gaseous globule (EGG). In the vicinity of massive stars, a prestellar core can be eroded by the ionizing radiation from massive stars, and may form a free-floating brown dwarf or planetary-mass object \citep{1996AJ....111.2349H,2004A&A...427..299W}. The final mass of the formed object can be approximated as $0.010~M_{\odot}~(\frac{a_{1}}{0.3~{\rm km~s}^{-1}})^{6}~(\frac{\dot{N}_{\rm Lyc}}{10^{50}~{\rm s}^{-1}})^{-1/3}~(\frac{n_{0}}{10^3~{\rm cm}^{-3}})^{-1/3}$ \citep{2004A&A...427..299W}, where $a_{1}$ is the isothermal sound speed in the neutral gas of the core, $\dot{N}_{\rm Lyc}$ is the emission rate of Lyman continuum photons, and $n_{0}$ is the number density of protons in the H\,{\scriptsize II} region around the core. In ONC, we take $\dot{N}_{\rm Lyc}\sim2.6\times10^{49}~s^{-1}$ \citep{1996ApJ...458..222B}, and $n_{0}\sim5.5\times10^{3}$ around Proplyd~133-353 \citep{2015A&A...582A.114W}. Assuming $a_{1}\sim0.23-0.36$\kms\ for a prestellar core with a gas temperature of 15--35\,K, the expected final mass of the object formed in the core near Proplyd~133-353 would be around 2--28\,$M_{\rm J}$, which is consistent with the mass (13\,$M_{\rm J}$) of Proplyd~133-353.}

\section{conclusion}

We have presented new near-infrared spectroscopy of Proplyd 133-353, a candidate photoevaporating disk given its cometary shape. The near-infrared spectroscopic data and  {\rev and the location on the H-R diagram} suggests that its central source is an M9.5-type substellar object with a mass ($\lesssim$ 13\,$M_{\rm J}$) close to the planetary regime, and an age younger than 0.5\,Myr. Proplyd~133-353 has been detected by Chandra, and presents a similar ratio of X-ray luminosity to stellar luminosity as other young stars in the ONC  with similar stellar luminosity. The comparison of two-epoch HST images suggests that Proplyd~133-353 has a similar proper motion to the one of the ONC. We explore two possibilities to explain the existence of Proplyd~133-353 near $\theta^{1}$\,Ori\,C, and propose that  Proplyd~133-353 was formed in a very low-mass dusty cloud or {\rev an evaporating gaseous globule} near $\theta^{1}$\,Ori\,C in a second-generation of star formation, which can explain both its young age and the estimated lifetime of its disk.

 The discovery of Proplyd~133-353 may provide a clue to understanding the significant peak at 10--20~$M_{\rm J}$ of the ONC initial mass function \citep{2002ApJ...573..366M,2016MNRAS.461.1734D}. The second-generation star formation in the very low-mass globulettes {\rev or in the evaporating gaseous globules} could over-populate the objects of the ONC within such a mass range.

\acknowledgments
 Many thanks to Dr. Carmen Ortiz Henley for helping improve the language of the manuscript, and the anonymous referee for comments that help to improve this paper. This material is based upon work supported by the National Aeronautics and Space Administration under Agreement No. NNX15AD94G for the program ``Earths in Other Solar Systems''. The results reported herein benefitted from collaborations and/or information exchange within NASA’s Nexus for Exoplanet System Science (NExSS) research coordination network sponsored by NASA's Science Mission Directorate. CFM gratefully acknowledges an ESA Research Fellowship. This research is  based on observations made with the NASA/ESA Hubble Space Telescope, and obtained from the Hubble Legacy Archive, which is a collaboration between the Space Telescope Science Institute (STScI/NASA), the Space Telescope European Coordinating Facility (ST-ECF/ESA), and the Canadian Astronomy Data Centre (CADC/NRC/CSA). 


\end{document}